\documentclass[a4paper,fleqn,usenatbib]{mnras}

\usepackage{newtxtext,newtxmath}
\usepackage[T1]{fontenc}
\usepackage{ae,aecompl}
\usepackage[english]{babel}
\usepackage{graphicx}	% Including figure files
\usepackage{subfigure}
\usepackage{amsmath}	% Advanced maths commands
\usepackage{amssymb}	% Extra maths symbols
\hypersetup{draft}
\usepackage{multirow}

\def\objfirst{\mbox{cLBV1}}
\def\objsecond{\mbox{cLBV2}}

\title[New LBV candidates in the NGC\,247]{New luminous blue variable candidates in the NGC\,247 galaxy\thanks{Based in part on data collected at Subaru Telescope, which is operated by the National Astronomical Observatory of Japan}}

\author[Y. Solovyeva et al.]{
Y.Solovyeva,$^{1}$\thanks{E-mail:solovyeva@sao.ru}
A. Vinokurov,$^{1}$
A. Sarkisyan,$^{1}$
K. Atapin,$^{5}$
S. Fabrika, $^{1,2}$
A.~F. Valeev, $^{1}$
\newauthor
A. Kniazev,$^{3,4,5}$
O. Sholukhova,$^{1}$
O. Maslennikova $^{1}$
\\
$^{1}$Special Astrophysical Observatory, Nizhnij Arkhyz, Russia\\
$^{2}$Kazan Federal University, Kremlevskaya 18, 420008 Kazan, Russia\\
$^{3}$South African Astronomical Observatory, PO Box 9, 7935 Observatory, Cape Town, South Africa\\
$^{4}$Southern African Large Telescope Foundation, PO Box 9, 7935 Observatory, Cape Town, South Africa\\
$^{5}$Sternberg Astronomical Institute, Lomonosov Moscow State University, Universitetskij Pr. 13, Moscow 119992, Russia\\}

\date{Accepted XXX. Received YYY; in original form ZZZ}

\pubyear{2019}

\begin{document}
\label{firstpage}
\pagerange{\pageref{firstpage}--\pageref{lastpage}}
\maketitle

\begin{abstract}

We search for LBV stars in galaxies outside the Local Group. Here we present a study of two bright $H\alpha$ sources in the NGC\,247 galaxy. Object j004703.27-204708.4 ($M_V=-9.08 \pm 0.15^m$) shows the spectral lines typical for well-studied LBV stars: broad and bright emission lines of hydrogen and helium \ion{He}{i} with P Cyg profiles, emission lines of iron \ion{Fe}{ii}, silicon \ion{Si}{ii}, nitrogen \ion{N}{ii} and carbon \ion{C}{ii}, forbidden iron [\ion{Fe}{ii}] and nitrogen [\ion{N}{ii}] lines. The variability of the object is $\Delta B = 0.74\pm0.09^m$ and $\Delta V = 0.88\pm0.09^m$, which makes it reliable LBV candidate. The star j004702.18-204739.93 ($M_V=-9.66 \pm 0.23^m$) shows many emission lines of iron \ion{Fe}{ii}, forbidden iron lines [\ion{Fe}{ii}], bright hydrogen lines with broad wings, and also forbidden lines of oxygen [\ion{O}{i}] and calcium [\ion{Ca}{ii}] formed in the circumstellar matter. The study of the light curve of this star also did not reveal significant variations in brightness ($\Delta V = 0.29\pm0.09^m$). We obtained estimates of interstellar absorption, the photosphere temperature, as well as bolometric magnitudes $M_\text{bol}=-10.5^{+0.5}_{-0.4}$ and $M_\text{bol}=-10.8^{+0.5}_{-0.6}$, which corresponds to bolometric luminosities $\log(L_\text{bol}/L_{\odot})=6.11^{+0.20}_{-0.16}$ and $6.24^{+0.20}_{-0.25}$ for j004703.27-204708.4 and j004702.18-204739.93 respectively. Thus, the object j004703.27-204708.4 remains a reliable LBV candidate, while the object j004702.18-204739.93 can be classified as B[e]-supergiant.

\end{abstract}


%%%%%%%%%%%%%%%%%%%%%%%%%%%%%%%%%%%%%%%%%%%%%%%%%%

%%%%%%%%%%%%%%%%% BODY OF PAPER %%%%%%%%%%%%%%%%%%

\section{Introduction}
Luminous blue variables (LBVs) are the rare type of massive ($M\gtrsim25 M\odot$, \citealt{Humphreys16}) post-main sequence stars that is characterised by high luminosity of $> 10^5 L\odot$ and significant spectral and photometric variability on different time-scales.

 An evolutionary status of LBV is still unclear.The classic view of LBVs is that they correspond to a very short transition phase from single massive O-type stars to Wolf-Rayet type stars \citep{Groh14}. The evolution of O-stars of different initial masses can look as follows \citep{Crowther07}: \\

$M \sim 25-40 M\odot$: O $\rightarrow$ LBV$/$RSG $\rightarrow$ WN(H-poor) $\rightarrow$ SN Ib;\\

$M\sim 40-75 M\odot$: O $\rightarrow$ LBV $\rightarrow$ WN(H-poor) $\rightarrow$ WC $\rightarrow$ SN Ic;\\

$M>75 M\odot$: O $\rightarrow$ WN(H-rich) $\rightarrow$ LBV $\rightarrow$ WN(H-poor) $\rightarrow$ WC $\rightarrow$ SN Ic;\\

Some studies \citep{Galyam09, Kotak06} have suggested that they can be direct progenitors of core-collapse supernovae. 
\cite{Groh13} showed that rotating massive stars with initial masses of 20-25 $M\odot$ can evolve according to the following scheme:

$20 M\odot$: O7.5V $\rightarrow$ BSG $\rightarrow$ RSG $\rightarrow$ BSG$/$BHG $\rightarrow$ LBV $\rightarrow$ SN;\\

$25 M\odot$: O6V $\rightarrow$ OSG $\rightarrow$ RSG $\rightarrow$ OSG$/$WNL $\rightarrow$ LBV $\rightarrow$ SN.\\

 \citep{Smith15} showed that LBVs may be the result of the evolution of close binaries, what is evidenced by their noticeable isolation from massive O-type stars compared to other massive stars (O star subtypes, WR stars and other classes of evolved  stars).  However, in work of \cite{Humphreys16} these results are called into question.

%TO_DO correct
The phenomenon of LBV is still not well understood, which is mainly associated with a rareness of known LBVs. Only LBVs and cLBVs in the galaxies of the Local Group were studied in detail  \citep{Richardson18}: at the time of 2018, only 41 LBVs and 108 cLBVs were discovered. However, data about the objects of this type in the Local Volume are scarce and only a few LBV stars are known beyond 1 Mpc (for example, \cite{Pustilnik16, Humphreys19, Drissen97, Goranskij2016NGC2770} and others). Consequently, the discovery of new LBVs would significantly help to understand the evolutionary status and interconnection of LBVs with other massive stars. In addition, an increase in the number of known LBVs will make it possible to find out whether LBVs deficiency with luminosity between $\log(L_\text{bol}/L_{\odot})$ 5.6 and 5.8 is real or is it the result of a small number of known LBVs \citep{Smith2004}.

The search for LBVs in Galaxy is complicated by the low accuracy of distance determination caused by strong dust absorption in the Galactic plane. Many LBVs are surrounded by compact circumstellar envelopes of different morphology \citep{Nota1995, Weis2001} originated during eruptions, and therefore are invisible in the optical range. Therefore, the most effective way to detect circumstellar shells is to use infrared telescopes (for example, Spitzer space Telescope, the Wide-field Infrared Survey Explorer). Over the past decade, several new LBVs and cLBVs in  Galaxy have been discovered using this method \citep{Gvaramadze2010a, Gvaramadze2010b, Gvaramadze2015, Kniazev2016}.  Unfortunately, distances to the Galactic LBVs are often not accurately determined even with the Gaia data. This leads to unreliable estimates of their luminosities.

The mostly accurate distance measurements and the slight interstellar absorptions make the nearby galaxies the ideal laboratory for studying of LBVs, which may helping to bridge the gap between theory and observations. One of the most commonly used searching method involves looking for $H\alpha$ emissions associated with blue stars \citep{Massey2006, Sholukhova97, Valeev10a, Neese, Corral}.

The variability of the LBV stars is irregular and often has a form of outbursts. The most extreme outbursts, or giant eruptions, having amplitudes of $>2.5^m$ are very rare; they are observed at times of hundreds-thousands of years. The stars that exhibit such a type of variability are called $\eta$ Car variables \citep{Humphreys99}. During the giant eruptions, the bolometric luminosity of the star does not stay constant, and a noticeably increase in the mass loss rate may occur which can lead to the formation of an emission nebula (as in the case of $\eta$ Carinae).

The more frequent outbursts with amplitudes from $0.1^m$ to $2.5^m $ (S Dor variables) are observed on time-scales of years to decades \citep{Genderen2001}. In such outbursts the bolometric luminosity remains approximately constant. The apparent brightening of the star occurs due to changes in the bolometric correction with decreasing star temperature. During the visual maximum, the LBV absorption spectrum is similar to spectra of A-F--type supergiants. When the visual brightness decreases, the LBV spectrum resembles that of an evolved hot supergiant of B-type or an Of/late--WN stars\citep{Vink12} and the stellar temperature can reach more than 35000 K \citep{Clark05}

%Fig.1
\begin{figure*} 
\centering 
\includegraphics[width=0.35\linewidth]{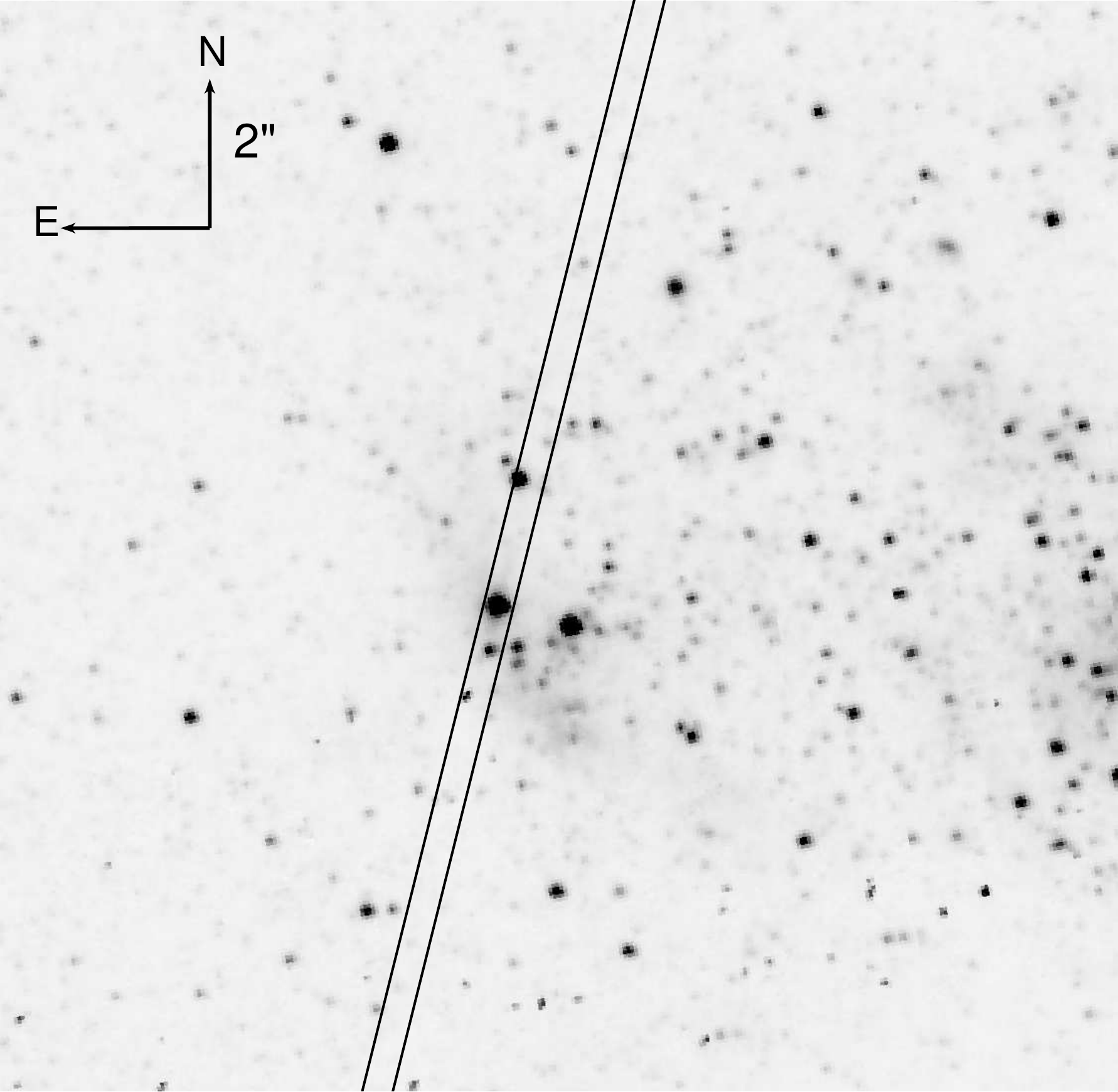}
\hspace{4ex}
\includegraphics[width=0.35\linewidth]{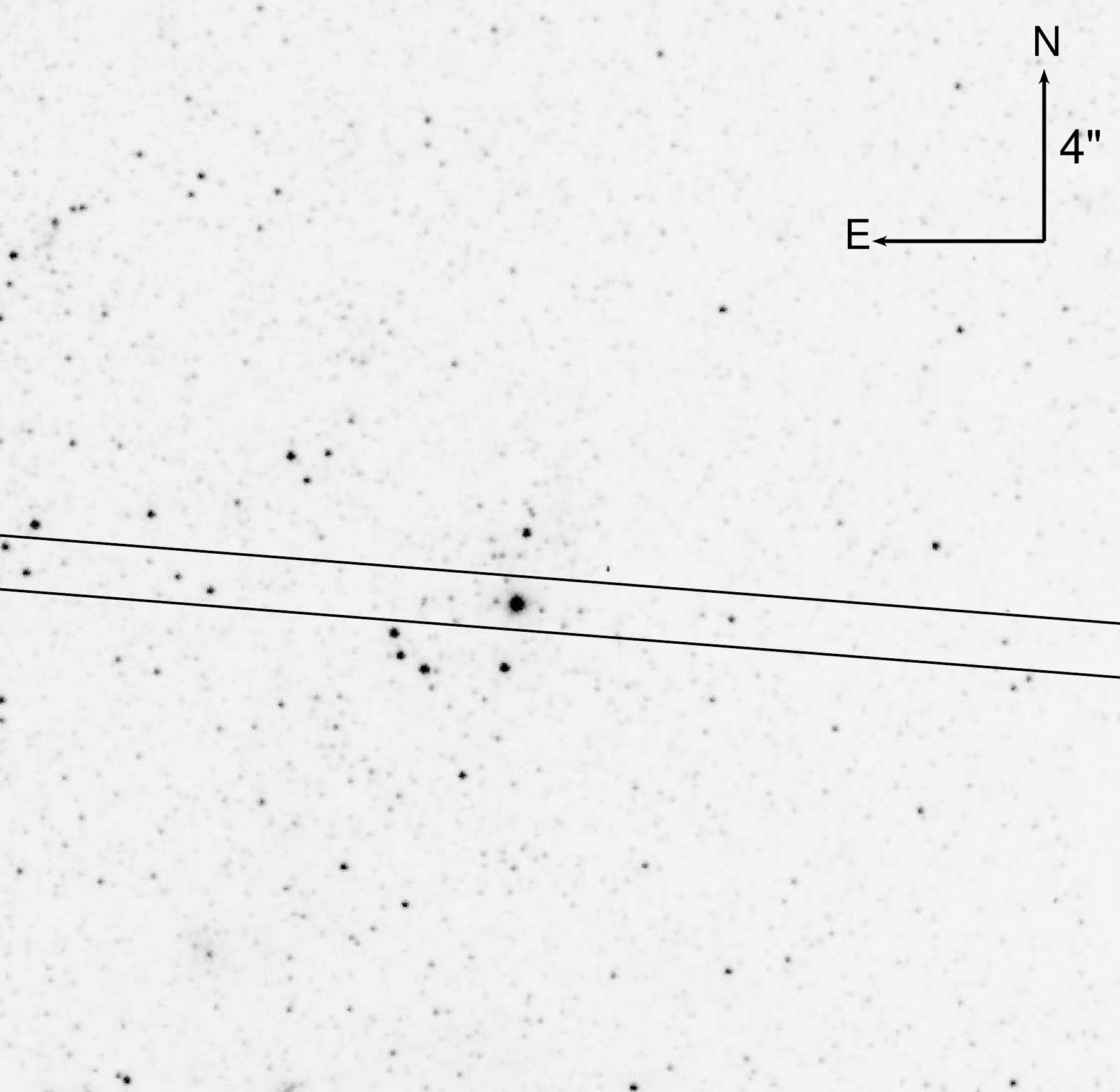} 
\caption{HST/ACS/F606W images of \objfirst\ ({\it left-hand panel}) and \objsecond\ ({\it right-hand panel}). The positions of the 0.4\arcsec\ slit used for the Subaru spectroscopy of \objfirst\ and the 1.25\arcsec slit\ for the SALT spectroscopy of \objsecond\ are also shown.} 
\label{Fig1} 
\end{figure*}

In some periods of their lives, LBVs show spectra very similar to those of B[e] supergiants (\citealt{Zickgraf06} and references therein) (sgB[e])~--- another type of bright ($4\leqslant\log(L_{bol}/L_{\odot})\leqslant6$) post-main sequence stars \citep{Kraus2005}. In contrast to LBV stars, sgB[e] are not so variable (about $0.1^m$--$0.2^m$) \citep{Lamers98} and demonstrate forbidden emission lines [\ion{O}{i}] $\lambda$ 6300,6364 \citep{Aret} and [\ion{Ca}{ii}] $\lambda$ 7291,7324 which indicate the presence of a dust envelope around the star. The two-component wind model with a hot polar wind and a slow and cold equatorial disk wind \citep{Zickgraf86, Fabrika2000} was proposed to explain their spectra. The spectral energy distributions (SEDs) of sgB[e] show significant near infrared excess which also supports the idea of the hot circumstellar dust envelope \citep{Zickgraf86,Bonanos09}.

\cite{Humphreys14} have divided all high-luminosity stars into six types according to their spectral and photometric features: Of/late-WN stars, LBVs, warm hypergiants, Fe\,II-emission stars, hot and intermediate supergiants. Despite the similarity of some observational features (spectra, SEDs) of these types of stars, the evolutionary connections between them have not yet been understood. Of all these types of the massive stars only LBVs have significant brightness variability, which make the detection of the S Dor-type variability the most important and classifying property.
 However, it is worth noting that this type of variability can also be observed in B-type supergiants\citep{Kalari18}, but it is not accompanied by spectral variability. \cite{Kalari18} concluded that B-supergiants and LBV stars are likely objects of different nature, and not all B-supergiants become S Doradus variables.

We search LBVs and similar objects in Local Volume galaxies using the archival broadband and near H$\alpha$ narrowband images, obtained with ACS, WFPC2 and WFC3 cameras of Hubble Space Telescope (HST). The new candidates which are showing H$\alpha$ emission of point-like blue stellar source were picked based on the following selection criterion: the source should be point-like and bright in all filters. 

So, our survey of Local Volume galaxies which aim to discover of LBVs, firstly presented in the paper \citep {Solovyeva19}, is continuing with study of  NGC\,247. The dwarf spiral galaxy (SAB(s)d) NGC\,247 is one of the closest spiral galaxies of the southern sky. The star formation rate and metallicity of this galaxy is similar to those of M33 galaxy, but NGC\,247 is more inclined towards the line of sight. The H$\alpha$ luminosity of NGC\,247 is comparable to those ones of M31, M33 and LMC \citep{Kennicutt2008}.

In this paper we present the discovery and detailed investigation of two new LBV candidates in NGC 247: j004703.27-204708.4 and j004702.18-204739.93 (hereafter \objfirst\ and \objsecond, respectively).

\section{Observations and data reduction}

In this work we used long-slit spectroscopic observations obtained with the Subaru (Hawaii) and the SALT (South Africa) telescopes as well as photometry from the HST, Spitzer and ground-based telescopes.

\subsection{Spectra}

We downloaded the Subaru data for \objfirst\ from the SMOKA Science Archive\footnote{\url{https://smoka.nao.ac.jp/}}. The observation was conducted on 2016 October~8 (proposal ID o16146) with the  Faint Object Camera and Spectrograph (FOCAS) \citep{Yoshida00, Kashikawa}. The 300B grating together with a 0.4\arcsec slit (its orientation is shown in Fig.~\ref{Fig1}) were used. The original spectral range was 3650--8300\,\AA\ and the resolution was 5\,\AA, however, due to contamination by scattered light at shorter wavelengths and order overlapping at the longer ones we decided to trim the range by 3800--7400\,\AA. Seeing was 0.4-0.6\arcsec, which allowed to separate \objfirst\ from the nearest bright star ($V=19.78^m\pm0.03^m$, $H\alpha=19.37^m\pm0.04^m$) located 1.0\arcsec\ away from the object (Fig.~\ref{Fig1}, left panel).

The SALT \citep{Buckley06, Donoghue06}  spectrum of \objfirst was obtained on 2017 November~11 with the RSS spectrograph \citep{Burgh03, Kobulnicky} using the PG0900 grating and a slit of 1.25\arcsec, which revealed the  spectral range 4300--7400\,\AA\ and resolution 5.3\AA. To suppress contribution from higher spectral orders, the PC03850 UV filter was used. Seeing turned out to be relatively bad $\simeq 1.7$\arcsec, which did not allow to resolve \objfirst\ and the nearest star.
However, after the data reduction, we did not find any changes in the emission lines compared to the spectrum obtained in 2016 with the Subaru telescope. Therefore, below we discuss only the spectrum obtained in 2016.

The spectrum of \objsecond\ was obtained on 2018 October~7 with the SALT telescope using the same equipment as in the case of the previous object. The seeing was 1.6\arcsec.

The Subaru spectrum was reduced with the context LONG of \textsc{midas} using the standard algorithm. A correction for the distortion was performed with the FOCASRED package\footnote{\url{https://www.naoj.org/Observing/DataReduction/Cookbooks/FOCAS_cookbook_2008dec08}} in the \textsc{iraf} environment. The primary processing of the SALT spectra were carried out with the PySALT package \citep{Crawford}, and further reduction was performed in \textsc{midas}. The spectra were extracted using the \textsc{spextra} package \citep{Sarkisyan2017} developed to deal with long-slit spectra in crowded stellar fields.

%Table 1: Optical photometry
\begin{table*}
\begin{minipage}{18cm}
\caption{Results of the optical photometry. The columns show the instruments, dates and  observed stellar magnitudes {\bf (not corrected for reddening)}.}
\begin{tabular}{lccccccc} \hline\hline 
 \centering
Telescope & Date  & U, mag &  B, mag & V, mag & R, mag &$H\alpha$, mag & I, mag \\ \hline
\multicolumn{8}{c}{\objfirst} \\ \hline
 2.2-m ESO/MPG/WFI &2002/10/08 & --- & $19.39\pm0.08$ &  $19.27\pm0.08$ & --- & --- &  --- \\
  2.2-m ESO/MPG/WFI& 2003/06/24& --- & $19.29\pm0.07$ & $19.17\pm0.08$ & --- & --- & ---\\
 2.2-m ESO/MPG/WFI & 2003/10/19& --- & ---& $19.17\pm0.07$ & --- & --- & --- \\
 2.2-m ESO/MPG/WFI & 2003/10/22& --- & $19.30\pm0.07$ &--- & --- & --- & --- \\
 ESO/VLT/FORS2 & 2004/06/19& --- & $19.38\pm0.05$ &$19.22\pm0.06$ & --- & --- & --- \\
 2.2-m ESO/MPG/WFI & 2004/10/17& --- & $19.28\pm0.09$ &$19.17\pm0.07$& --- & --- & --- \\
  2.2-m ESO/MPG/WFI & 2005/01/06& --- & --- &$19.10\pm0.09$& --- & --- & --- \\
 2.2-m ESO/MPG/WFI & 2005/06/09 & --- & --- &$18.83\pm0.08$& --- & --- & --- \\
 2.2-m ESO/MPG/WFI & 2005/09/26 & --- & $19.15\pm0.08$ &$19.01\pm0.09$& --- & --- & --- \\
 2.2-m ESO/MPG/WFI & 2006/09/25 & --- & $19.30\pm0.08$ &$19.22\pm0.10$& --- & --- & --- \\
HST/ACS/WFC& 2011/10/11 & --- & $19.89\pm0.04$ & $19.71\pm0.04$ & --- & $17.34\pm0.03$ & --- \\
Subaru & 2016/10/09$^{*}$ & $18.83\pm0.10$ & $19.54\pm0.11$ & $19.45\pm0.09 $ & $19.13\pm0.09 $ & --- & $19.10\pm0.09 $\\
2.5-m, SAI MSU  & 2018/09/20 & --- &  --- & $19.37\pm0.10$ & $19.01\pm0.10$ & ---\\ \hline

\multicolumn{8}{c}{\objsecond} \\ \hline
JKT & 1994/08/16 & --- & $18.94\pm0.09$ & --- & $18.81\pm0.10$ & --- & ---  \\
Danish 1.54-m & 1997/09/08 & $18.27 \pm 0.12$ & $18.82 \pm 0.07$ & $18.89 \pm 0.07$ & $18.90\pm 0.11$& --- & ---   \\  2.2-m ESO/MPG/WFI &  2002/10/08 & --- & $19.08\pm0.07$ &  $18.96\pm0.08$ & --- & --- & --- \\
  2.2-m ESO/MPG/WFI & 2003/06/24 & --- &  $19.01\pm0.06$ & $18.94\pm0.07$ & --- & --- & --- \\
 2.2-m ESO/MPG/WFI & 2003/10/19 & --- & ---& $18.88\pm0.07$ & --- & --- & --- \\
 2.2-m ESO/MPG/WFI & 2003/10/22 & --- & $18.96\pm0.07$ &--- & --- & --- & --- \\
 ESO/VLT/FORS2 & 2004/06/19 & --- & $19.02\pm0.05$ &$18.96\pm0.06$ & --- & --- & --- \\
 2.2-m ESO/MPG/WFI & 2004/10/17 & --- & $19.09\pm0.07$ &$18.98\pm0.06$& --- & --- & --- \\
 2.2-m ESO/MPG/WFI &  2005/06/09& --- & --- &$18.94\pm0.08$& --- & --- & --- \\
 2.2-m ESO/MPG/WFI &2005/09/26 & --- & $18.95\pm0.06$ &$18.87\pm0.08$& --- & --- & --- \\
 2.2-m ESO/MPG/WFI & 2006/09/25& --- & $18.98\pm0.07$ &---& --- & --- & --- \\
CTIO 0.9m & 2009/08/21 & $18.16\pm0.13$ & $18.82\pm0.10$ & $18.88\pm0.11$ & $18.87\pm0.11$ & --- & --- \\ 
HST/ACS/WFC& 2011/10/11 & --- & $18.83\pm0.03$ & $18.71\pm0.04$ & --- & $17.25\pm0.03$ & --- \\ 
Subaru & 2016/10/09 & $18.38\pm0.10$ & $18.97\pm0.10$ & $18.97\pm0.09 $ & $18.78\pm0.09$ & --- & $18.63\pm0.09 $\\
2.5-m, SAI MSU  & 2018/09/20$^{*}$ & --- & --- & $19.00\pm0.08$ & $18.88\pm0.09$ & --- & ---\\ 
Zeiss-1000, SAO RAS & 2019/11/15 & --- & $19.03\pm0.16$ & $18.95\pm0.10$ & $18.87\pm0.11$ & --- & ---\\ \hline

\end{tabular}
\label{Tab1}
\end{minipage}
\textit{Notes.}
$^{*}$ These photometric observations are simultaneous or quasi-simultaneous with spectroscopy. 
\end{table*}

\subsection{Imaging}

Images of the region containing both LBV candidates were obtained by the Subaru telescope together with the spectroscopic data. Additionally we observed the objects with the 2.5-m telescope of the Caucasian Mountain Observatory of SAI~MSU and the Zeiss-1000 of SAO~RAS, and also involved archival data from  other telescopes: the Danish 1.54-m telescope, the Cerro Tololo Inter-American Observatory (CTIO) 0.9-m telescope, the Jacobus Kapteyn Telescope (JKT). We also using the archival data from Very Large Telescope (VLT) and 2.2-m MPG telescope of ESO. Only images with seeing of $\approx 0.8\arcsec$ or better were chosen, where \objfirst\ and neighbour star are resolved.
A summary of the utilised observations is given in Table~\ref{Tab1}. Primary data reduction was performed with \textsc{midas}.  
We carried out an aperture photometry  with the APPHOT package of \textsc{iraf} with applying aperture correction to measure the observed magnitudes and using 13 nearby stars as reference for absolute calibration. The aperture sizes were from 0.3 to 1.2\arcsec\ depending on seeing in a particular observation. The background level was measured within source-free ring apertures, the size of which was selected depending on seeing.

To perform an absolute photometric calibration we used archival HST observations. Each of the objects was observed once in the optical band (with the ACS/WFC camera, the F435W, F606W and F658N filters) and once in the IR band (the WFC3/IR camera, the F105W and F160W filters). Optical photometry was carried out using a $0.15$\arcsec\ circular aperture. Such a small aperture size was chosen to avoid contamination from nearby sources. The background was determined in a ring aperture with an inner radius of $0.25$\arcsec\ and a thickness of $0.2$\arcsec. The aperture correction was taken into account by photometric measurements of 20 single bright stars in a large (1\arcsec) and small (0.15\arcsec) apertures.

The measured HST magnitudes were converted into the standard Johnson-Cousins system with PySynphot package assuming spectral indices $-1.2$ for \objfirst\ and $-1.7$ for \objsecond\  obtained from spectroscopy. The F435W filter passband is close to B, while the passband of the F606W filter is between V and R, and its peak is close to the $H\alpha$ line. Therefore, to prevent overestimation of the obtained V-band magnitudes due to the presence of bright $H\alpha$ emission lines in the spectra of our objects, we estimated the contribution of the $H\alpha$ emissions using the available spectral data and subtracted them from the observed F606W fluxes. All the optical magnitudes are shown in Table~\ref{Tab1}.

The IR source fluxes were measured using archival data of WFC3/IR camera of HST in a $0.32$\arcsec circular aperture, the background level in $1.3 - 2.6$ \arcsec annulus. Aperture corrections were calculated by measuring the fluxes of 10 single bright stars in $0.32$\arcsec\ and $0.42$\arcsec\ apertures. Results are presented in Table~\ref{Tab2}.

Also, we used archival data obtained with the Spitzer telescope in the $3.6\mu$m and $4.5\mu$m bands of the IRAC camera \footnote{\url{https://sha.ipac.caltech.edu/applications/Spitzer/SHA)}}. Observations in the other IRAC filters we did not analyse due to their low spatial resolution. We performed absolute photometry of the objects using a 0.9\arcsec\ aperture.The choice of aperture size was due to the crowded field of stars. The background level was determined in a ring aperture with an inner radius of $1.5$\arcsec and a thickness of $0.9$\arcsec. Aperture corrections were calculated from photometry of 25 single bright stars in the large (6\arcsec) and small apertures (0.9\arcsec). Then we compared our results for single stars with the photometry from the Spitzer point source catalogue \citep{Khan15}. The results have coincided within the errors for the 25 reference stars and for \objsecond, while the fluxes of \objfirst\ (together with the nearest star that Spitzer cannot resolved) in the $3.6\mu$m and $4.5\mu$m filters appeared to be lower by 25 and 15 percents, respectively, due to a more accurate accounting for the background  in our work. Extrapolating the spectral slope between the F105w and F160W band flux to the range of $3.6 \mu$m and $ 4.5 \mu$m, we estimated the contribution of the nearest star to the total flux as $\lesssim10$\%.  The flux measured in 0.9$\arcsec$ aperture includes a significant part of the flux of the relatively compact nebula which surrounding \objfirst\ and part of nearby stellar association, the value of which we cannot reliably estimate. The obtained Spitzer magnitudes are also shown in Table~\ref{Tab2}.

%Table 2: IR  vertical version
\begin{table}
\hspace{-3mm}
\begin{minipage}{8.5cm}
\caption{Results of the IR photometry.} 
\begin{tabular}{ccccc} \hline\hline 
 & \multicolumn{2}{c}{HST/WFC3/IR} & \multicolumn{2}{c}{Spitzer/IRAC} \\
 & \multicolumn{2}{c}{(2014/06/30)} & \multicolumn{2}{c}{(2012/10/17)} \vspace{1ex} \\
Object & F105W, & F160W, & $3.6\mu$m, & $4.5\mu$m, \\ 
& mag & mag & mag & mag \\ \hline
\objfirst & $19.17 \pm 0.04$ & $18.57 \pm 0.04$ & $16.20 \pm 0.16$ & $16.09\pm0.14$ \\
\objsecond & $18.67 \pm 0.04$ & $18.13 \pm 0.04$ & $15.91 \pm 0.13$ & $15.25\pm0.11$ \\ \hline
\end{tabular} 
\label{Tab2}
\end{minipage}
\end{table} 

\section{Results}
%Fig. 2
\begin{figure*} 
\centering 
\includegraphics[angle=270,scale=0.7]{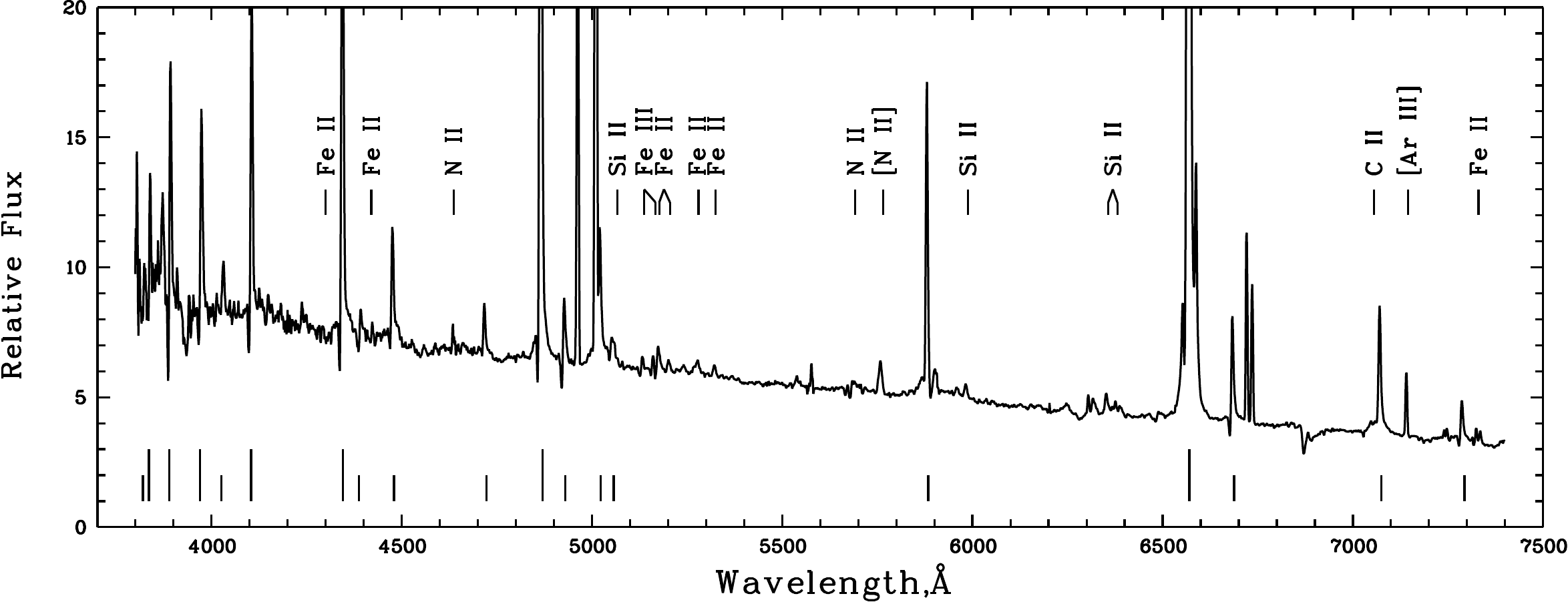}\\
\vspace*{2ex}
\includegraphics[angle=270,scale=0.7]{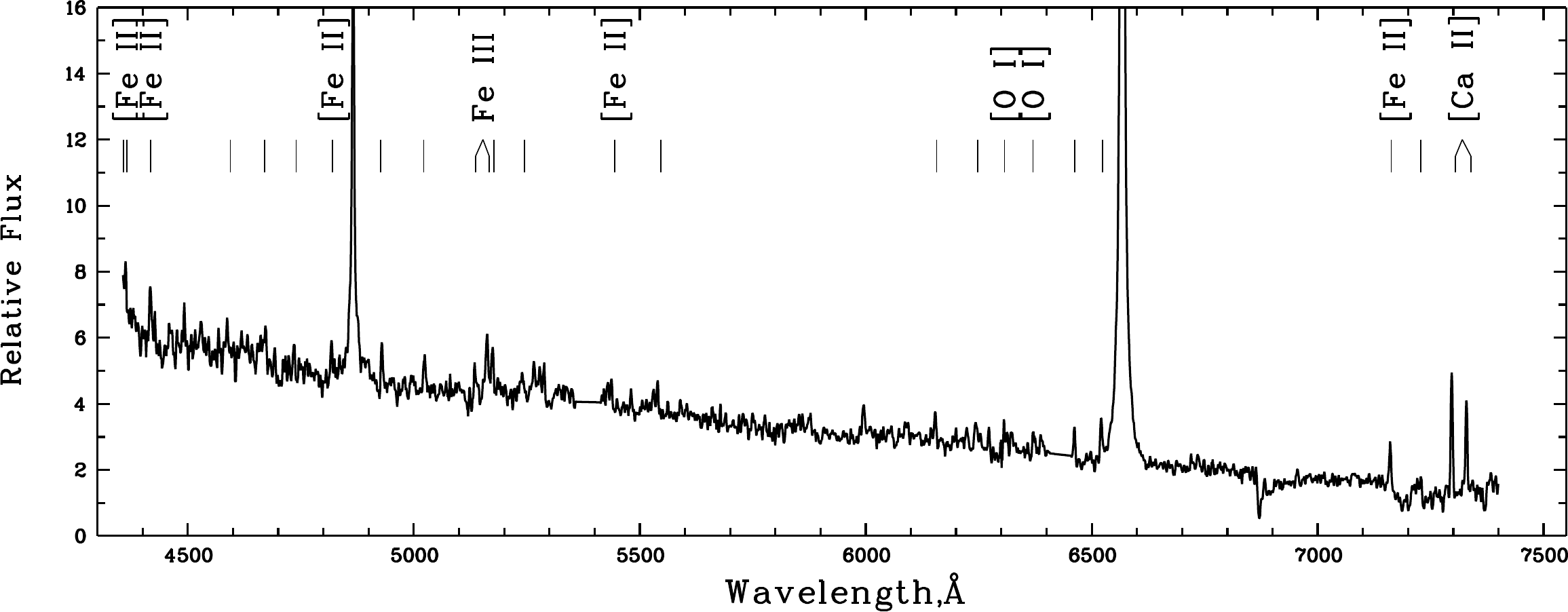}
\caption{{\it Top panel:} spectrum of \objfirst\ obtained with the Subaru telescope. The unlabelled long and short ticks designate the hydrogen Balmer lines and [\ion{He}{i}] lines, respectively. Narrow components of H$\alpha$, H$\beta$, H$\gamma$ and H$\delta$ as well as forbidden lines [\ion{O}{i}], [\ion{N}{ii}] and [\ion{S}{ii}] belong to the surrounding nebulae. {\it Bottom panel:} spectrum of \objsecond\ obtained with SALT. The unlabelled  ticks designate the \ion{Fe}{ii} lines.}
\label{Fig2} 
\end{figure*}

%Fig. 3
\begin{figure*}
\includegraphics[angle=0, width=0.46\linewidth]{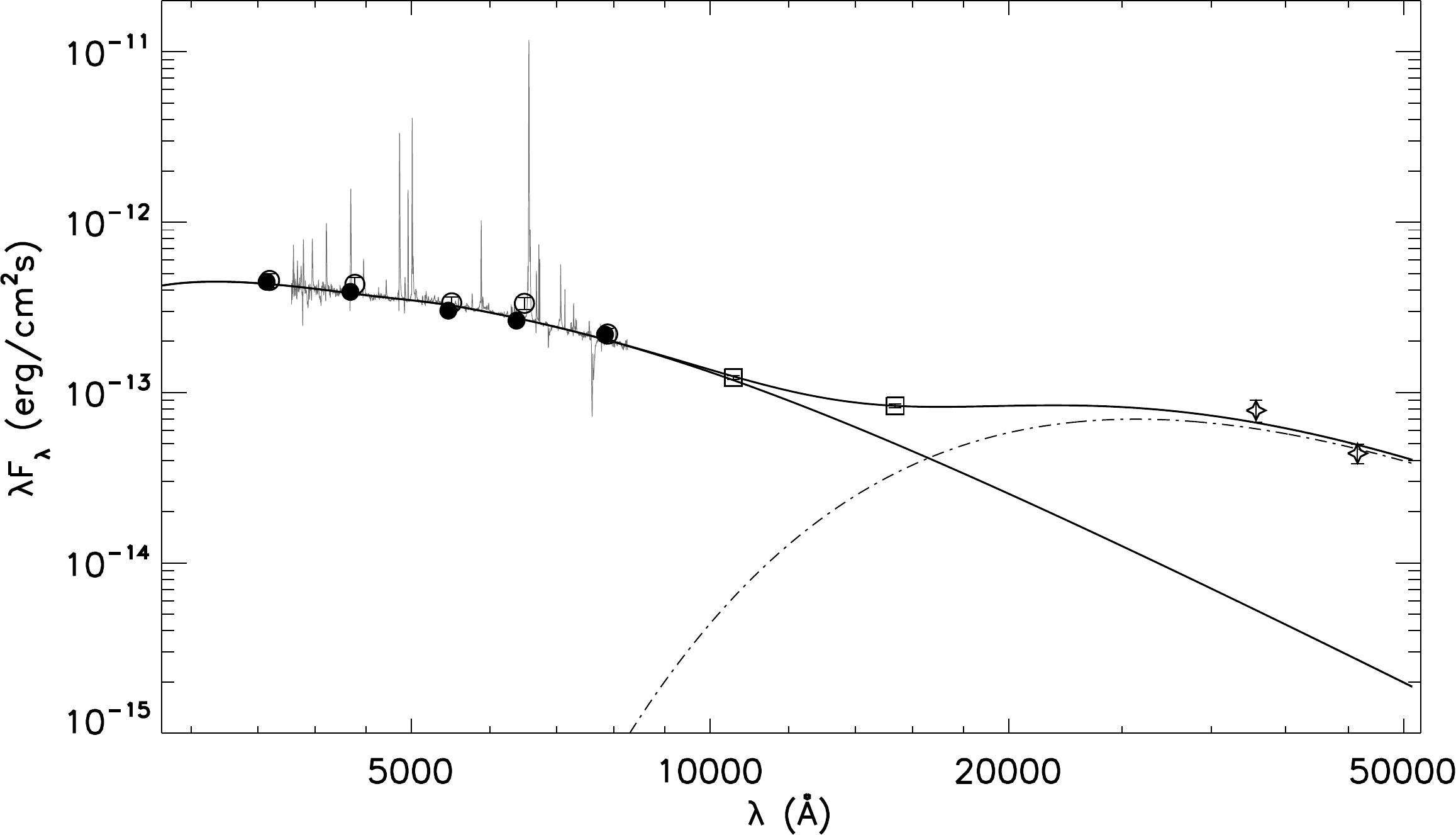}
\includegraphics[angle=0, width=0.46\linewidth]{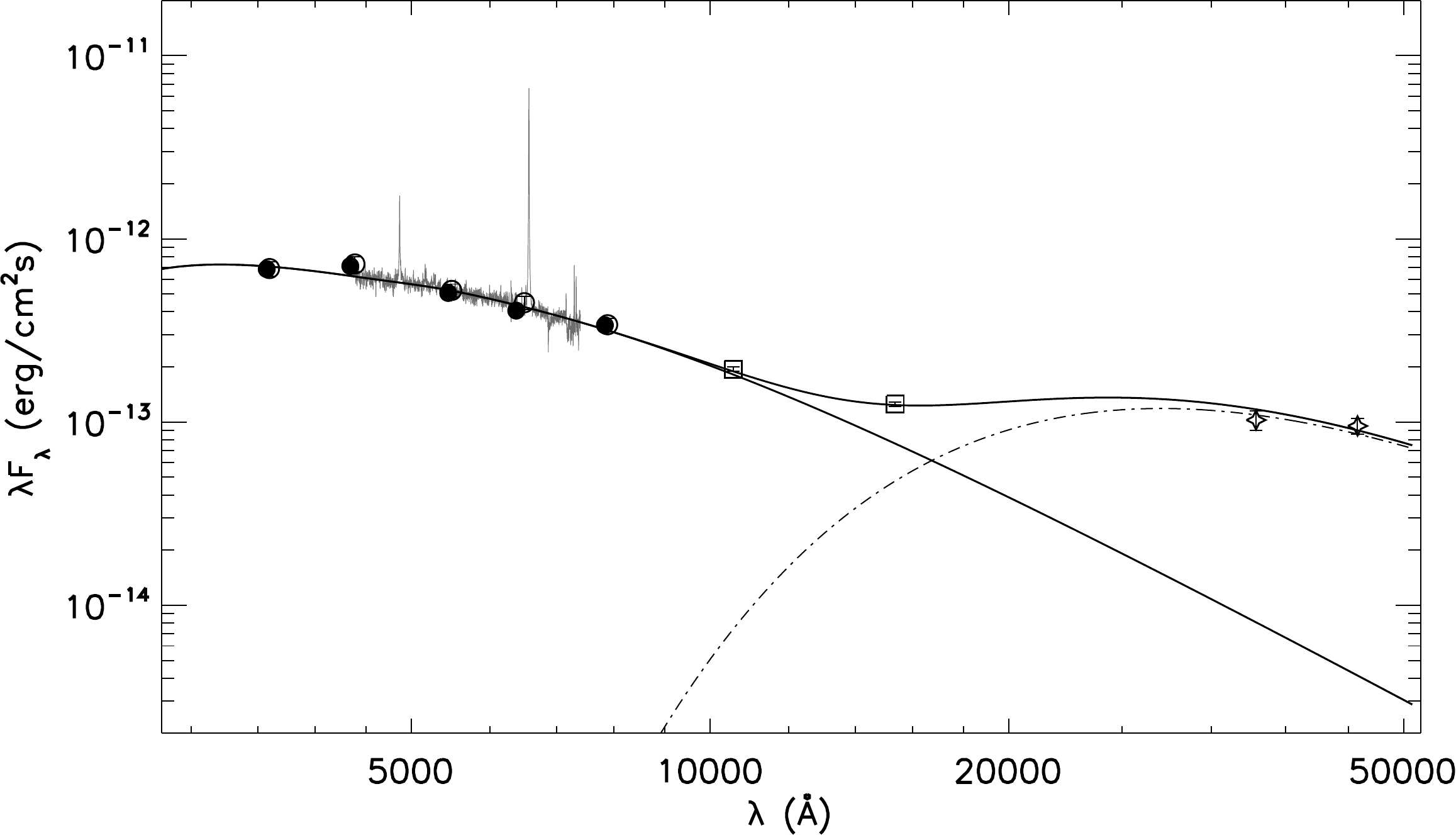}  
\caption{Spectral energy distibutions (SEDs) of \objfirst\ ({\it left-hand panel}) and \objsecond\ ({\it right-hand panel}). The circles indicate the $UBVR_cI_c$ photometry from Subaru (2016), filled circles denotes the fluxes corrected for the contribution of bright emission lines, unfilled circles~--- original fluxes. The squares and crosses indicate the IR photometry obtained with HST/WFC3/IR in the F105W and F160W bands (2014) and Spitzer/IRAC in the $3.6\mu$m and $4.5 \mu$m band (2012), respectively. Observed spectra the same as in Fig.~\ref{Fig2} are shown by grey. The SEDs were fitted by the model (solid line) consisting of two black body components describing emission from the stellar photosphere (dashed line) and the IR excess (dash-dotted line) arising probably due to the dust envelope around the star. 
 The best-fitting model parameters are $T_\text{SED}=18000 \pm 2000$\,K and $15000\pm2000$\,K, $A_V\approx 0.9$ and $0.7$, $T_\text{dust}\approx 1400$\,K and  1300\,K for \objfirst\ and \objsecond, respectively.}
\label{Fig3} 
\end{figure*}

%Table 3
\begin{table*}
\caption{Parameters of the studied stars. The columns show the object name, the reddening measured from the nebular hydrogen lines, the photosphere temperature estimated from both the 
spectra and SEDs, the absolute V-band and bolometric magnitudes and the bolometric luminosity.} \begin{tabular}{ccccccc} \hline\hline
Star & $A_V$,~mag  &  $T_\text{spec}$,~K & $T_\text{SED}$,~K & $M_V$,~mag & $M_\text{bol}$,~mag &  $\log(L_\text{bol}/L_{\odot})$  \\ \hline
\objfirst & $0.80 \pm 0.10$ & $20000 \pm 5000$ & $18000 \pm 2000$ & $-9.08 \pm 0.15$ & $-10.5^{+0.5}_{-0.4}$ & $6.11^{+0.20}_{-0.16}$  \\ \hline
\objsecond & $0.90 \pm 0.20$ & $15000 \pm 5000$ & $15000 \pm 2000 $ & $-9.66 \pm 0.23$ & $-10.8^{+0.5}_{-0.6}$ & $6.24^{+0.20}_{-0.25}$\\ \hline
\end{tabular} 
\label{Tab3}
\end{table*}

\subsection{\objfirst\ (j004703.27-204708.4)}

The Subaru spectrum of \objfirst\ is presented in Fig.~\ref{Fig2} (top panel). It shows broad and strong emission lines of hydrogen and helium \ion{He}{i} with P Cyg profiles. Also, there are many lines of iron \ion{Fe}{ii}, \ion{Fe}{iii}, silicon \ion{Si}{ii} and weak emission lines of nitrogen  \ion{N}{ii} $\lambda$4631 and carbon \ion{C}{ii} $\lambda$7053. The presence of these lines indicates that the photosphere temperature should be about $T_\text{spec} = 20000 \pm 5000$~K. Using the full width at half-maximum (FWHM) of the forbidden nitrogen line [\ion{N}{ii}] $\lambda$5755 forming in the stellar wind \citep{Stahl1991, Crowther1995}, we derived the terminal velocity $V_{\infty}=464\pm26$~km\,s$^{-1}$. Other bright forbidden lines [\ion{O}{iii}], [\ion{Ar}{iii}], [\ion{N}{ii}] $\lambda$6548, $\lambda$6583, [\ion{S}{ii}] $\lambda$6717, $\lambda$6731 are probably emitted from the surrounding nebula. 
Based on the ratio of the hydrogen lines in the surrounding nebula, we estimated the reddening as $A_V = 0.80 \pm 0.10^m$ assuming  the case B photoionization \citep{Osterbrock}.

The temperature was also measured from the photometric data.
We have constructed a spectral energy distribution (SED) of the object (Fig.~\ref{Fig3}a) combining the $UBVR_cI_c$ photometry from Subaru with the IR photometry obtained with the {\it HST} and {\it Spitzer} telescopes (however, we should note that the IR data are not simultaneous with the optical ones). The photometric data shown in the figure are corrected for the contribution of the emission lines. However, since the spectrum does not completely cover the filter U range, the contribution of emission lines to the flux in this filter might be underestimated. The SED was fitted by the Planck function taking into account interstellar absorption with $R_V = 3.07$ \citep{Fitzpatrick}. Possible variations of the temperature and reddening were restricted to vary within uncertainties obtained from the spectroscopy $T_\text{spec}=20000\pm5000$~K and $A_V=0.80\pm0.10^m$. Eventually we have obtained the best-fitting temperature of $T_\text{SED} = 18000 \pm 2000$~K for $A_{V} \approx 0.9$.
Also we have found an evidence of significant IR excess probably caused by the presence of a circumstellar dust envelope. We fitted this excess by adding an extra black body component with temperature $T_{\text{dust}}\approx1400$~K (Fig.~\ref{Fig3}a).

Based on the SED best-fit temperature $T_\text{spec} = 18000 \pm 2000$~K and the reddening $A_V=0.80\pm0.10^m$ obtained from spectroscopy, we estimated the absolute V-band and bolometric magnitudes as $M_V=-9.08 \pm 0.15$~mag and $M_\text{bol}=-10.5^{+0.5}_{-0.4}$~mag, which corresponds to the bolometric luminosity of $\log(L_\text{bol}/ L_{\odot})=6.11^{+0.20}_{-0.16}$. Both derived magnitudes are quite typical for LBV stars \citep{Humphreys94}.

\begin{figure} 
\includegraphics[angle=270,scale=0.3]{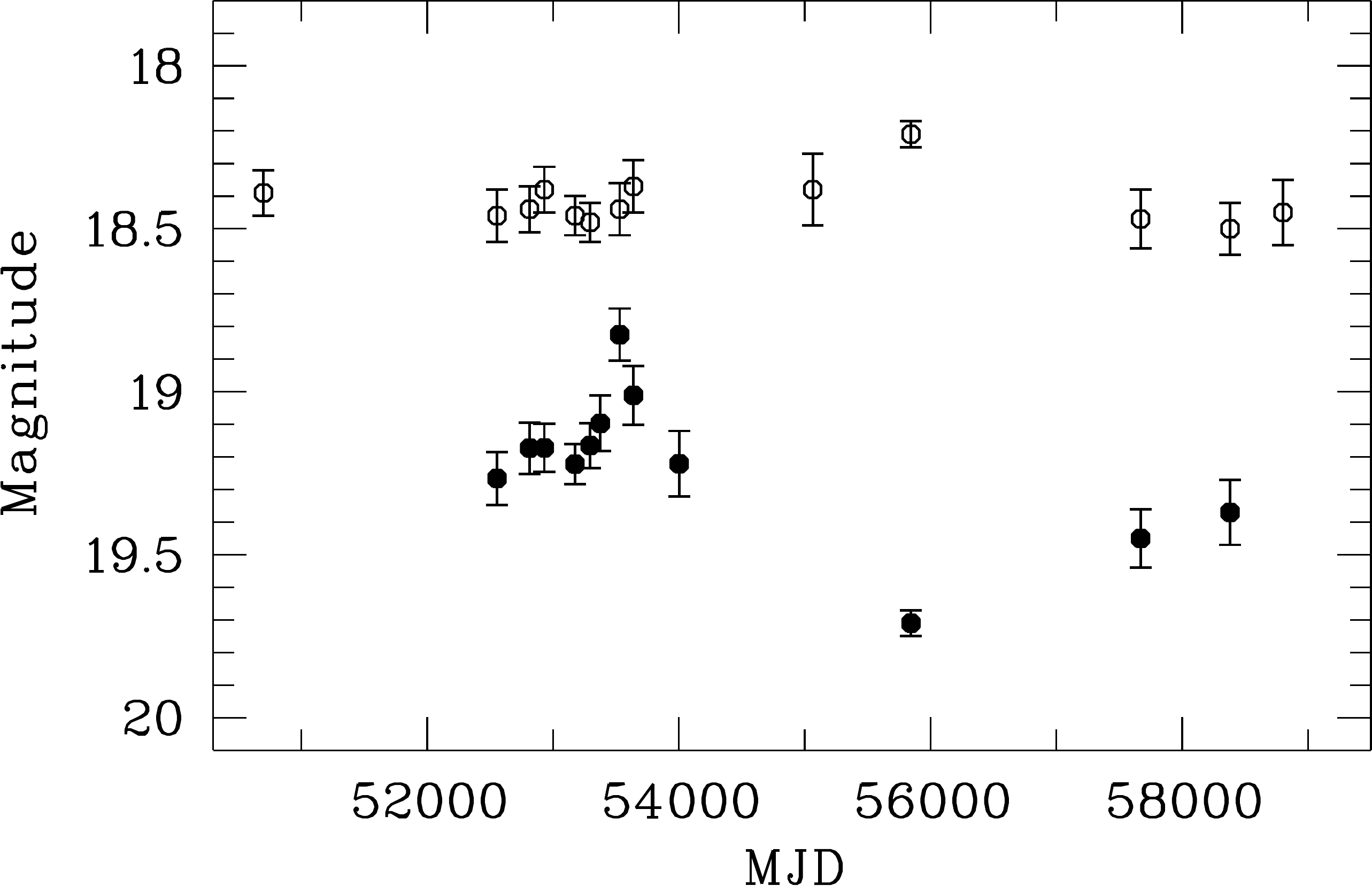}  
\caption{ Light curves of \objsecond (open symbols) and \objfirst (filled symbols) in V band. To better visualisation light curve of \objsecond was shifted on -0.5$^m$. } 
\label{Fig4} 
\end{figure}

 The object \objfirst\ demonstrated significant variability from 2005 (2.2-m ESO/MPG/WFI) to 2011 (HST/ACS/WFC) (Table~\ref{Tab1}): the source B and V-band magnitudes changed by $0.74 \pm 0.09^m$ and $0.88 \pm 0.09^m$ respectively. Such a strong change in brightness makes this object a reliable candidate for bona fide LBV. The light curve of this object are shown on Figure \ref{Fig4}. Further observations of the object are necessary to study the nature of the photometric variability.

 %Fig with cLBVs light curves 
 
%Fig with temperature-luminosity diagram
\begin{figure} 
\includegraphics[angle=0,scale=0.55]{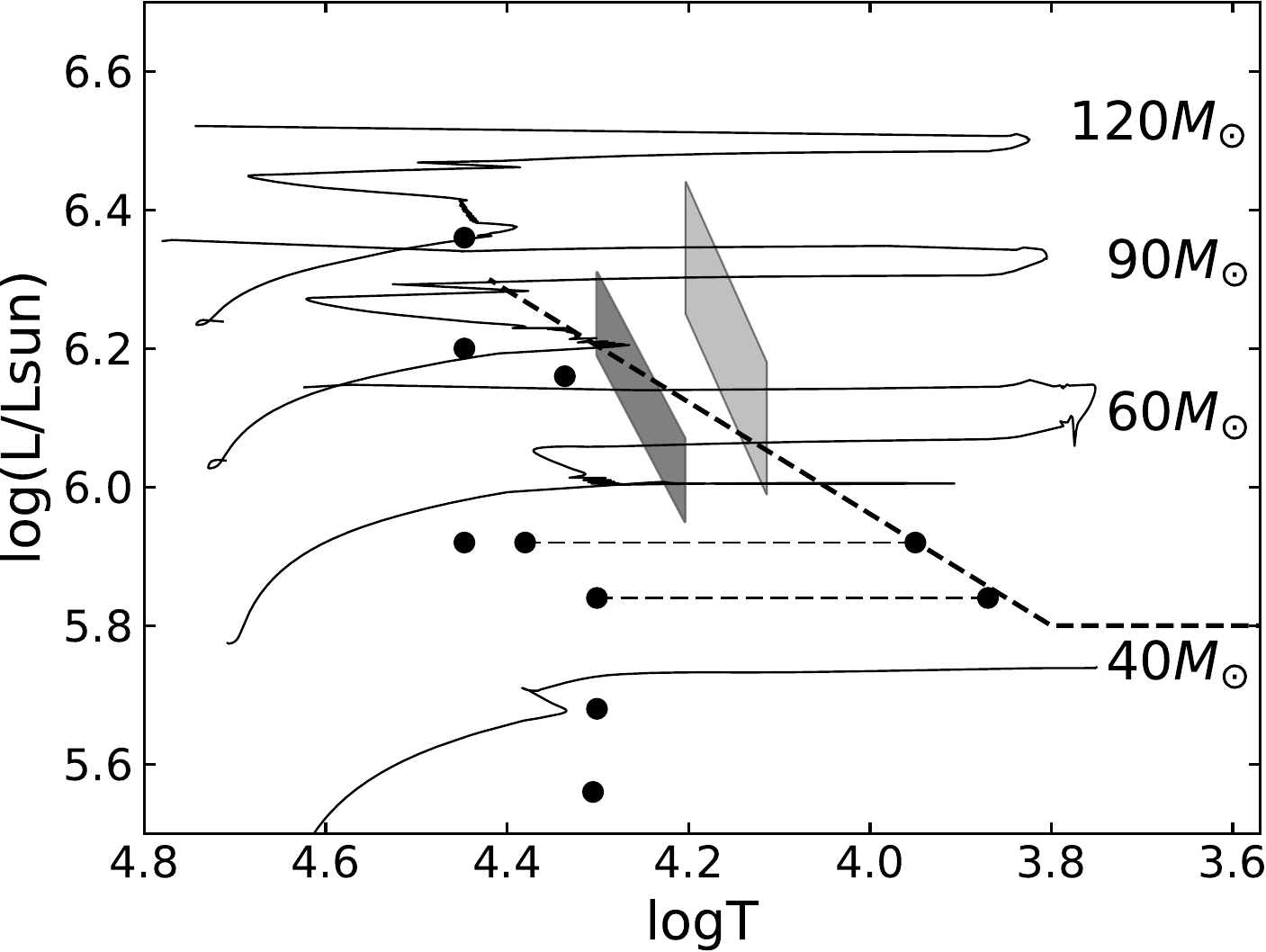}  
\caption{Temperature-luminosity diagram with evolutionary tracks of massive stars for the metallicity of Z = 0.008 \citep{Tang}. The possible positions of \objfirst and \objsecond are shown by dark grey and light grey areas, respectively. Here we used the temperature ranges obtained from the SED modelling. A thick dash line indicates the Humphreys-Davidson limit. For illustrative purposes, we have shown by black circles the position of the confirmed LBV stars from M31 and M33 galaxies and a thin dash lines denote LBV transits for Var B and Var C (taken from \citealt{Humphreys17b}).} 
\label{Fig5} 
\end{figure}

\subsection{\objsecond\ (j004702.18-204739.93)}

The spectrum of \objsecond\ contains broad emission Balmer lines $H\alpha$, $H\beta$ as well as a large number of iron lines \ion{Fe}{ii} and \ion{Fe}{iii}. The multiple forbidden iron lines [\ion{Fe}{ii}] indicate the presence of the powerful outflow of the stellar matter. The observed lines \ion{Fe}{ii} have a maximum intensity when the temperature is about $15000$ K, and weaken if the temperatures decrease below $10000$ or increase above $20000$~K. Therefore, we estimate the photosphere temperature of \objsecond\ as $T_\text{spec} = 15000 \pm 5000$ K. The reddening measured from the ratio of the hydrogen lines in the surrounding nebula is $A_V = 0.90 \pm 0.20^m $.

The SED of \objsecond, as in the case of \objfirst, was constructed using the photometric data from the Subaru telescope together with the infrared observations from {\it HST} and {\it Spitzer}. The star did not show any brightness variations from 2016 to 2018, so we have used the spectrum from the SALT telescope obtained in 2018 to take into account the contribution of emission lines to the photometric data obtained with Subaru in 2016. The SED was approximated by the Planck function with the best fit temperature $T_\text{SED} = 15000 \pm 2000 $~K.  
As in the case of the previous object, the SED of \objsecond\ also shows an IR excess, which may be produced by the dust envelope. We estimated the temperature of the dust component as  $T_{\text{dust}} \approx 1300$~K.

Using the photosphere temperature $T_\text{SED} = 15000 \pm 2000$~K and the reddening estimated from the nebular hydrogen lines $A_V=0.90 \pm 0.20$~mag we determined the absolute and bolometric magnitudes $M_V=-9.66 \pm 0.23$~mag and $M_{bol}=-10.8^{+0.5}_{-0.6}$~mag, which corresponds to the bolometric luminosity $\log(L_{bol}/L_{\odot})=6.24^{+0.20}_{-0.25}$.

 The photometric variability of \objsecond\ is $\Delta V \approx 0.29 \pm 0.09^m$ from 2011 (HST/ACS/WFC) to 2018 (2.5-m, SAI MSU), which only slightly exceeds the 3$\sigma$ level. The light curve of this object are shown on Fig.\ref{Fig4}.

\section{Discussion and conclusions}

Above we have measured the stellar parameters of our LBV candidates residing in the galaxy NGC\,247 (Table~\ref{Tab3}) and found them very typical for LBV stars.
In Fig.~\ref{Fig5} we show the temperature--luminosity diagram overplotted with evolutionary tracks of massive stars. Since there are no spectroscopic estimates for the metallicity of this galaxy, and different authors use values from $Z=0.004$ \citep{Wagner-Kaiser} to $Z=0.0152$ \citep{Rodrigez}, we assumed the metallicity of $Z=0.008 $ (as in \citealt{Davidge}) to plot the evolutionary tracks \citep{Tang}. Grey areas show ranges of possible photosphere temperatures and luminosities of the candidates. 
As can be seen in the figure, both \objfirst\ and \objsecond\ have to be more massive then 25 $M_{\odot}$ that is enough for a star to pass the LBV stage. Our objects are located near the area of LBV stars in the temperature--luminosity diagram. 
%Fig. 6,7

\begin{figure*}
\includegraphics[angle=0, width=0.46\linewidth]{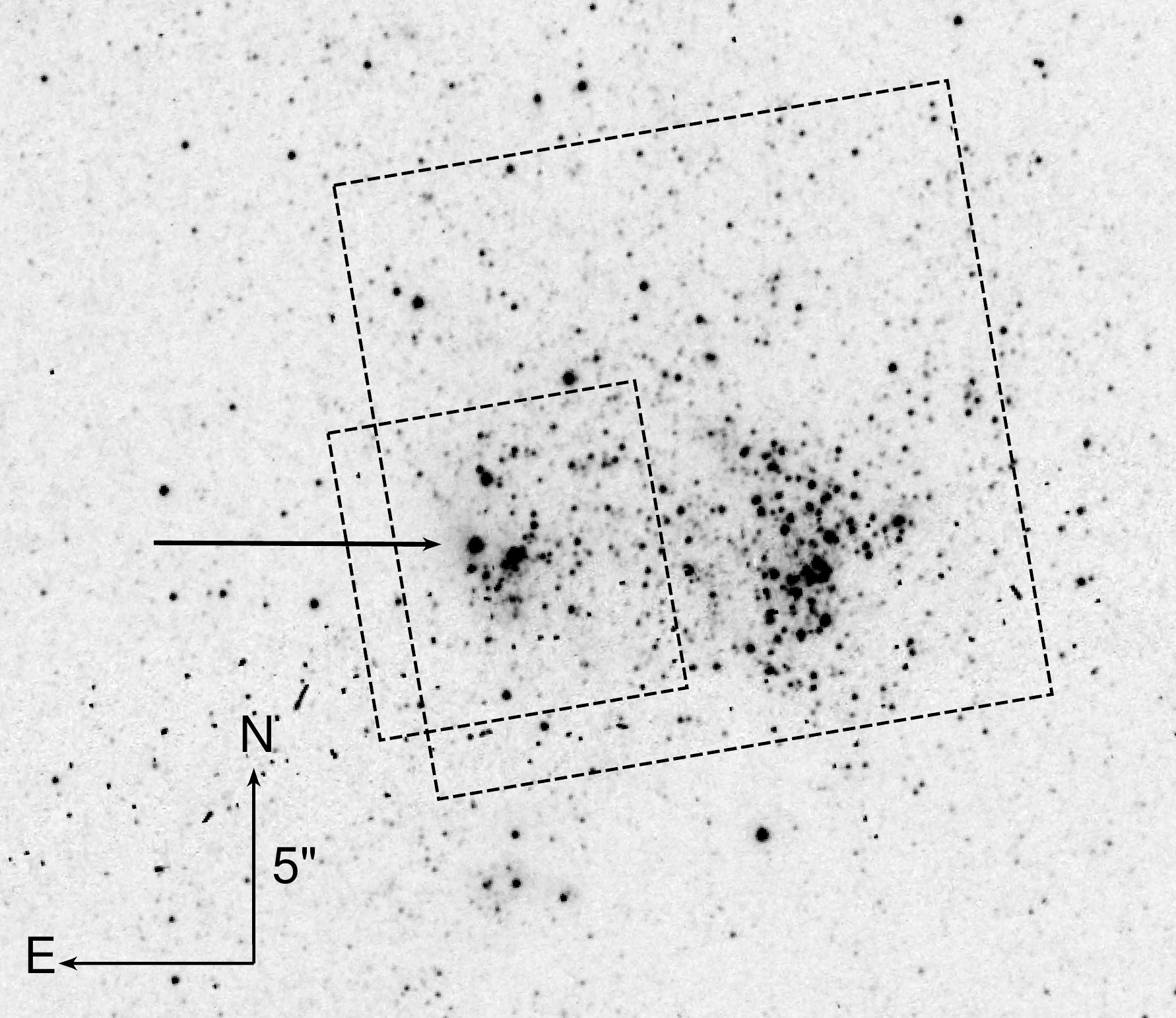}
\includegraphics[angle=0, width=0.46\linewidth]{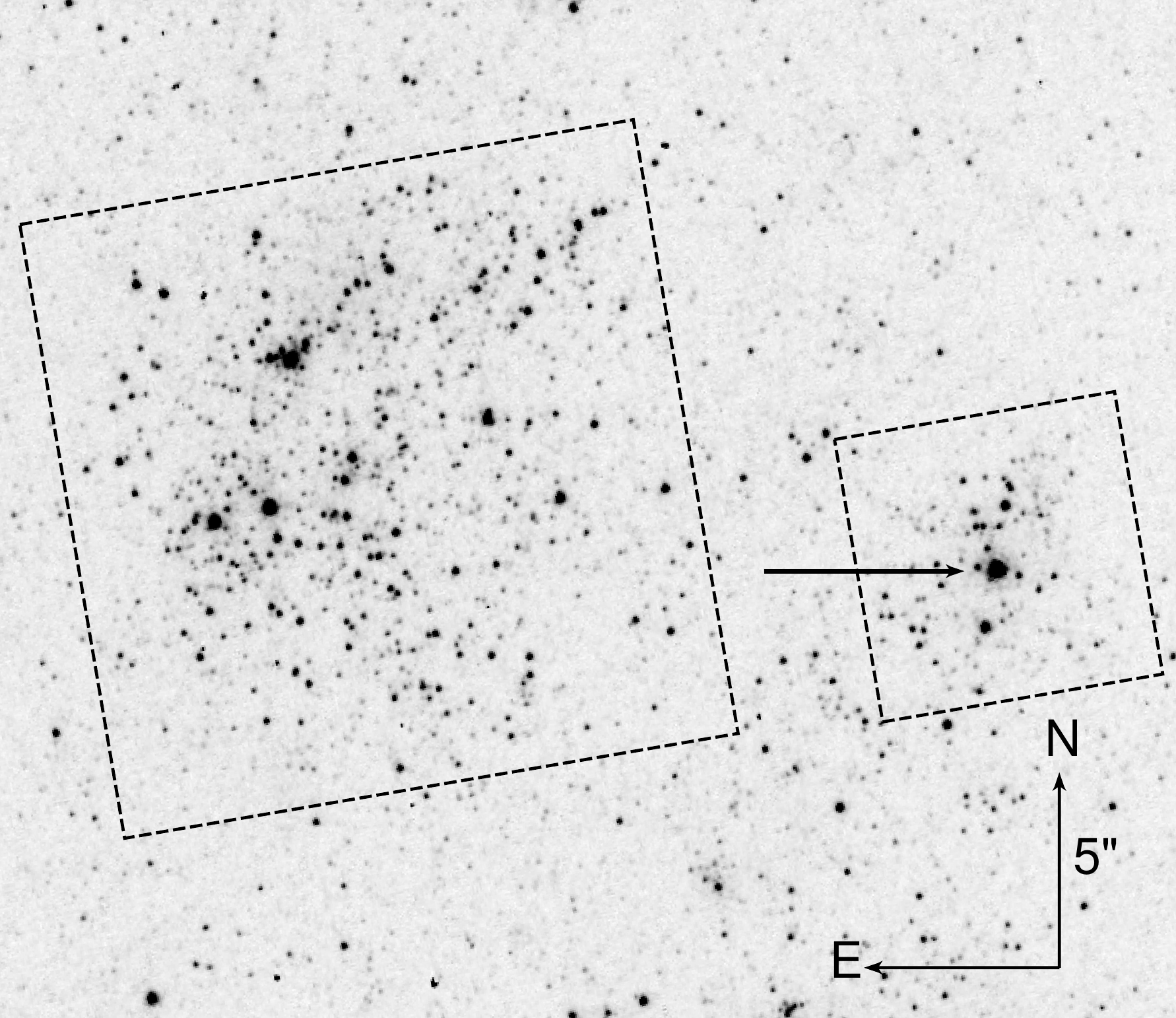}  
\caption{ HST/ACS/WFC image of \objfirst (left) and \objsecond (right) in F435W band. Studied cLBVs are marked with arrows. The squares designate stellar associations with size of 8\arcsec $\times$ 8\arcsec and 16\arcsec $\times$ 16\arcsec that were used to construct the colour-magnitude diagrams.}
\label{Fig6} 
\end{figure*}

\begin{figure*}
\includegraphics[angle=0, width=0.46\linewidth]{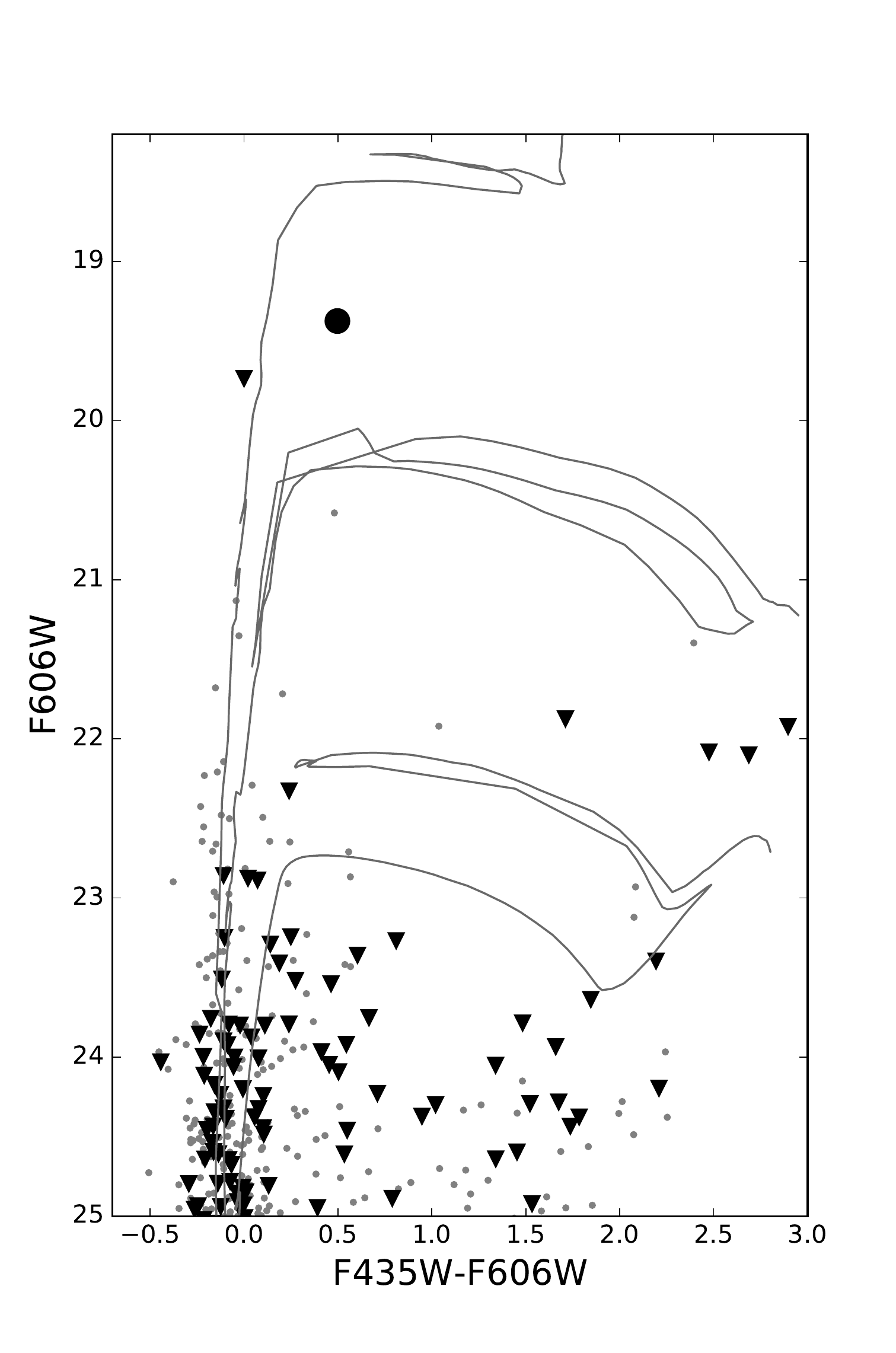}
\includegraphics[angle=0, width=0.46\linewidth]{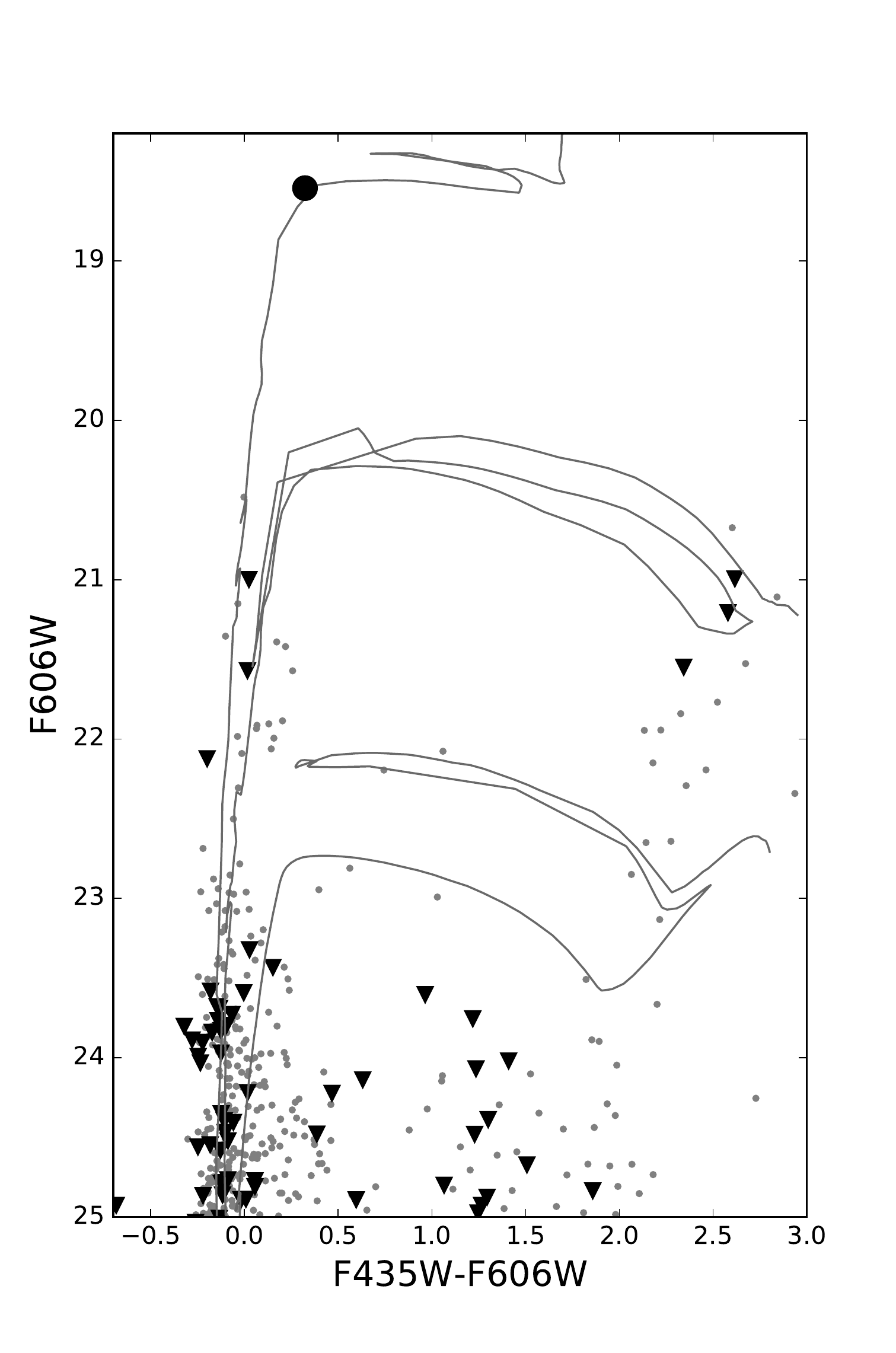}  
\caption{ Color-magnitude diagrams for stellar associations close to \objfirst (left) and \objsecond (right). The grey dots mark large (16\arcsec $\times$ 16\arcsec) area, the black triangles denote small area (8\arcsec $\times$ 8\arcsec) of stars, black circles - cLBVs. Theoretical isochrones \citep{Marigo17} for metallicity Z = 0.008 of 5, 10 and 30 Myr are shown (from top to bottom).}
\label{Fig7} 
\end{figure*}

 We have found that \objfirst\ shows brightness variations of $0.74 \pm 0.09^m$ and $0.88 \pm 0.09^m$ in B and V bands respectively, which noticeably exceed the $3\sigma$ level. Moreover, its spectrum is similar to those of the Galactic bona fide LBV WS\,1 \citep{Kniazev15} obtained in 2011. However, the spectral energy distribution of \objfirst\ shows a noticeable near-IR excess: we obtained flux ratio of $F_{V}/F_{3.6{\mu}m} \approx 60$, while confirmed LBVs have $F_{V}/F_{3.6{\mu}m} \textgreater 100$ (e. g., \citealt{Humphreys14}). Thus, the IR excess of \objfirst\ is 2 times higher than the values observed in confirmed LBVs. However, it may be due to the underestimation of the contribution of the neighboring star and the nebula surrounding the \objfirst. Moreover, we also do not observe spectral variability, although we have only two spectra obtained with a time interval of 1 years (2016 and 2017). Therefore, this object remains in the LBV candidate status.

The second object \objsecond\ shows the forbidden emission lines [\ion{Ca}{ii}] $\lambda\lambda$\,7291,7324 and [\ion{O}{i}] $\lambda\lambda$\,6300,6364, which are assumed to be indicators of the innermost disc regions in sgB[e] \citep{Kraus2007, Aret2012}. These lines are also observed in warm hypergiants \citep{Humphreys2013}. In addition, \objsecond\ shows a noticeable IR-excess, which is characteristic of stars of these types. Due to the lack of absorption features in its spectrum, the object cannot be attributed to warm hypergiants. So, taking into account the presence of forbidden iron lines [\ion{Fe}{ii}] along with forbidden lines of oxygen and calcium, we could classify \objsecond\ as a B[e] supergiant (\citealt{Humphreys17} and references therein).

 Both studied stars are close to stellar associations, which are likely to be parental to them.
There are relatively small groups of stars around the cLBVs, as well as rich star-forming regions at some distances (see Figure \ref{Fig6}). The sizes of the areas where are located most of the blue stars are 8\arcsec $\times$ 8\arcsec and 16\arcsec $\times$ 16\arcsec. To estimate the ages of these groups we have performed PSF photometry of thr stars and have constructed the "color-magnitude" diagrams (CMDs).
 
Photometry for selected regions was made using the DOLPHOT and archival images, obtained on F435W and F606W bands of HST/ACS/WFC. The CMDs were constructed, taking into account a mean extinction $E(B-V)=0.18$ in NGC\,247 \citep{Gieren2009} (Figure \ref{Fig7}). Theoretical isochrones from work of \citet{Marigo17} for metallicity $Z = 0.008$ and cLBVs were also plotted on the diagram.

Comparison with theoretical isochrones showed the possible continuous star formation in the studied stellar associations. However, the position of majority of the stars in the CMDs is well described by the isochrones of 10-30 Myr.
 It is worth noting that both cLBVs lie on the diagram noticeably higher than most of the stars (excluding the bright star nearby \objsecond). This can be explained by the younger age of the candidates in the case of the single star evolution. However, due to the relatively small number of stars in the studied associations, the upper part of the diagram may be poorly populated. Therefore, we can overestimate the age of the host star-forming region: the age of the last outburst of star formation can be much smaller and correspond to the age of the candidates. This assumption is confirmed by the bright star mentioned above, which is not cLBV, but is in the same region of the diagram as the candidates.

The position of the studied cLBVs in the CMDs can be explained in an alternative way. If we assume that both stars are the result of the evolution of close binaries with mass exchange, and the observed objects may be "rejuvenated" due to the accretion \cite{Beasor19, Smith15}, then this can explain the apparent isolation of the candidates from the stars of the nearest associations in the observed CMDs. 

It is worth noting that the question of the observed age of the studied objects and neighboring star-forming region requires a more detailed study, which is planned in the future.

Finally, we note that for the more accurate classification of \objfirst\ and \objsecond, one needs additional observations in order to search for more prominent photometric and spectral variability.

\section*{Acknowledgements}
This research was supported by the Russian Foundation for Basic Research 19-52-18007, 19-02-00432. The authors acknowledge partial support from M.V.Lomonosov Moscow State University Program of Development. Based on data obtained from the ESO Science Archive Facility under request number Yusoloveva/543153, Yusoloveva/543857, Yusoloveva/543876. Based on observations made with the NASA/ESA Hubble Space Telescope, obtained from the data archive at the Space Telescope Science Institute. STScI is operated by the Association of Universities for Research in Astronomy, Inc. under NASA contract NAS 5-26555. This work is based in part on observations (archival data) made with the Spitzer Space Telescope, which is operated by the Jet Propulsion Laboratory, California Institute of Technology under a contract with NASA. Some spectral observations reported in this paper were obtained with the Southern African Large Telescope (SALT) under program 2017-1-MLT-003. A.\,Kniazev acknowledges support from the National Research Foundation (NRF) of South Africa. Observations with telescopes of the SAO RAS are carried out with the support of the Ministry of Science and Higher Education of the Russian Federation (including agreement No05.619.21.0016, unique project identifier RFMEFI61919X0016).

\section*{Data Availability}
The data underlying this article will be shared on reasonable request to the corresponding author.

%%%%%%%%%%%%%%%%%%%%%%%%%%%%%%%%%%%%%%%%%%%%%%%%%%



\bibliographystyle{mnras} \bibliography{bibtexbase.bib}

%\begin{thebibliography}{99}
%\end{thebibliography}

%%%%%%%%%%%%%%%%%%%%%%%%%%%%%%%%%%%%%%%%%%%%%%%%%%

%%%%%%%%%%%%%%%%% APPENDICES %%%%%%%%%%%%%%%%%%%%%

% \appendix

% \section{Some extra material}

% If you want to present additional material which would interrupt the flow of the main paper,
% it can be placed in an Appendix which appears after the list of references.

%%%%%%%%%%%%%%%%%%%%%%%%%%%%%%%%%%%%%%%%%%%%%%%%%%

% Don't change these lines
\bsp	\label{lastpage} \end{document}